\documentclass[twocolumn,a4paper,showpacs,preprintnumbers,amsmath,amssymb,prl]{revtex4}

\usepackage{graphicx} 
\usepackage{dcolumn}
\usepackage{bm} 
\usepackage{color}

\begin{document}

\hyphenation{te-tra-go-nal}

\bibliographystyle{apsrev}

\title{Large spectral weight transfer in optical conductivity of SrTiO$_{3}$ induced by intrinsic vacancies}

\author{Teguh C. Asmara$^{1,2}$, Xiao Wang$^{1}$, Iman Santoso$^{1,2}$, Qinfang Zhang$^{3,4}$, Tomonori Shirakawa$^{3,4}$, Dongchen Qi$^{1,2}$, Aleksei Kotlov$^{5}$, Mallikarjuna R. Motapothula$^{1,6}$, Mark H. Breese$^{1,2}$, T. Venkatesan$^{1}$, Seiji Yunoki$^{1,3,4,7}$, Michael R\"{u}bhausen$^{1,8}$, Ariando$^{1}$, Andrivo Rusydi$^{1,2,3,8}$ }

\affiliation{$^{1}$NUSNNI-NanoCore, Department of Physics, National University of Singapore, Singapore 117576}
\affiliation{$^{2}$Singapore Synchrotron Light Source, National University of Singapore, Singapore 117603}
\affiliation{$^{3}$Computational Condensed Matter Physics Laboratory, RIKEN ASI, Wako, Saitama, 351-0198, Japan}
\affiliation{$^{4}$CREST, Japan Science and Technology Agency, Kawaguchi, Saitama 332-0012, Japan}
\affiliation{$^{5}$Hamburger Synchrotronstrahlungslabor (HASYLAB) at Deutsches Elektronen-Synchrotron (DESY), Notkestra$\beta$e 85, 22603 Hamburg, Germany}
\affiliation{$^{6}$Centre for Ion Beam Applications, Department of Physics, National University of Singapore, Singapore 117576}
\affiliation{$^{7}$Computational Materials Science Research Team, RIKEN AICS, Kobe, Hyogo 650-0047, Japan}
\affiliation{$^{8}$Institut f\"{u}r Angewandte Physik, Universit\"{a}t Hamburg, Jungiusstrasse 11, 20355 Hamburg, Germany. Center for Free Electron Laser Science (CFEL), D-22607 Hamburg, Germany}

\begin{abstract}

The optical conductivity ($\sigma_{1}$) of SrTiO$_{3}$ for various vacancies has been systematically studied using a combination of ultraviolet - vacuum ultraviolet (UV-VUV) reflectivity and spectroscopic ellipsometry. For cation (Ti) vacancies, $\sigma_{1}$ shows large spectral weight transfer over a wide range of energy from as high as 35 eV to as low as 0.5 eV and the presence of mid-gap states, suggesting that strong correlations play an important role. While for anion (O) vacancies $\sigma_{1}$ shows changes from 7.4 eV up to 35 eV. These unexpected results can be explained in terms of orbital reconstruction.

\end{abstract}

\maketitle

Intrinsic vacancies and orbital reconstruction are believed to be responsible for various exotic electronic quantum properties in complex oxide materials. An ideal example is SrTiO$_{3}$. The crystal structure of SrTiO$_{3}$ belongs to the perovskite family with cubic $Pm$\={3}$m$ space group at room temperature and a lattice parameter of 3.905 {\AA}, undergoing a symmetry-lowering transition to a tetragonal structure below 105 K \cite{Goodenough}. Optical and electrical conductivity studies reveal that SrTiO$_{3}$ is an insulator with a bandgap of 3.2 eV at room temperature \cite{LeePRB01,FrederiksePR}. It is inherently a quantum paraelectric, and with the aid of an external mechanical stress, such as from epitaxial strain of a thin film, it can achieve a stable ferroelectric phase \cite{UwePRB,MullerPRB,HaeniNature}. Strikingly, it can even exhibit superconductivity below 0.3 K induced by oxygen vacancies \cite{SchooleyPRL,CalvaniPRB,OhtomoJAP} and it shows a presence of unusually narrow ($<$ 2 meV) Drude peak at 7 K when doped with Nb \cite{vanMechelen08}.

Despite of many of the above-mentioned fascinating properties, roles of the intrinsic vacancies and orbital reconstruction in SrTiO$_{3}$ are still far from understood. One main problem is the limited number of reliable experiments to probe directly the electronic band structure in a broad energy range. Previous theoretical studies \cite{Eskes91,Ohta91,Meinder93} specifically said that a direct fingerprint of correlation effects (reflecting the orbital reconstruction) is the presence of spectral weight transfer in abroad energy range. Thus, a combination of ultraviolet - vacuum ultraviolet (UV-VUV) reflectivity and spectroscopic ellipsometry, which lead to a stabilized Kramers-Kronig transformation, is the most direct experiment to measure the complex dielectric response systematically in a broad energy range from which spectral weight transfer information can be obtained\cite{RusydiPRB, Santoso2011}.

Previous experimental studies either focused on reflectivity of pristine SrTiO$_{3}$ \cite{CardonaPR,vanBenthemJAP}, or on various modified forms of SrTiO$_{3}$ but in the photon energy region only near or much below the optical band gap \cite{FrederiksePR,vanMechelen08,MoosJACS}. In this Letter, we reveal the impact of intrinsic vacancies on the optical conductivity ($\sigma_{1}$) of SrTiO$_{3}$ by using the recently developed experimental method:  a combination of UV-VUV reflectivity (3.7 - 32.5 eV) and spectroscopic ellipsometry (0.5 - 6.5 eV) which cover a very wide photon energy range from 0.5 to 35 eV \cite{RusydiPRB,Santoso2011}. The complex dielectric function $\varepsilon(\omega)=\varepsilon_{1}(\omega)+i\varepsilon_{2}(\omega)$ from 0.5 to 6.5 eV is directly obtained from the self-normalized spectroscopic ellipsometry, which is absolute and independent from geometrical misalignments. The reflectivity spectra for this energy range is obtained from $\varepsilon(\omega)$ by using $R=\frac{|1-\sqrt{\varepsilon}|}{|1+\sqrt{\varepsilon}|}$, which is then used to normalized the UV-VUV reflectivity taking the advantage of the overlapping energy. The calibration of the UV-VUV reflectivity from 3.7 to 32.5 eV is done by comparing it with the luminescence yield of gold mesh and sodium salicylate (NaC$_{7}$H$_{5}$O$_{3}$). Furthermore, the UV-VUV reflectivity above 30 eV is normalized by using off-resonance scattering considerations\cite{HenkeADNDT}. We have shown that this procedure yields normalized reflectivity spectra in an energy range between 0.5 and 32.5 eV with a precision better than 0.3 percent\cite{RusydiPRB}. Details of the optical conductivity measurements are shown in online supplementary material \cite{supplementary}. We have repeated our experiments three times for each samples and the results are consistent.

The samples have been prepared by isochronal annealing of several pristine SrTiO$_{3}$ obtained from Crystec (99.99 percent purity) at 950 $^{\circ}$C for 30 minutes under a constant flow of oxygen at various ambient pressures ($P_{O_{2}}$): 5$\times$10$^{-7}$ (sample A), 1$\times$10$^{-5}$ (sample B), 1$\times$10$^{-3}$ (sample C), and 1$\times$10$^{-2}$ Torr (sample D). Prior to the optical measurements, the samples are characterized using X-ray diffraction spectrometer (XRD) and X-ray photoemission spectroscopy (XPS) to confirm their crystal structure and composition, respectively. The crystal structure of the samples remains unaffected by the annealing, while their composition is slightly altered. Based on our estimation and supported by first-principles plane-wave based calculations of various defect levels of SrTiO$_{3}$ (see supplementary\cite{supplementary}), O vacancies are found in sample A with concentration of 0.02 percent, while cationic vacancies are found on sample D with concentration of 0.01 percent. These are consistent with previous reports\cite{FrederiksePR,MoosJACS,TanakaPRB}. Noting that based on the diffusion coefficient of O at these temperatures \cite{PaladinoJPCS}, the depth affected by these vacancies ($\sim$10 $\mu$m) is significantly beyond the photon penetration depth (10 - 200 nm above the bandgap (see supplementary\cite{supplementary}). This relatively deep photon penetration depth also ensures that the experimental method is indeed bulk-sensitive, thus minimizing any surface conditions effects of a few angstroms on the measured optical spectra. Variable-angle spectroscopic ellipsometry measurements are also performed to further ensure this bulk-sensitivity and the samples's isotropy. Details of sample preparation, characterizations, and optical measurements are explained in more depths in the supplementary\cite{supplementary}.

Herewith, from the optical measurements, we find that $\sigma_{1}$ of SrTiO$_{3}$ varies as function of $P_{O_{2}}$ below and above the optical band gap up to 35 eV. Surprisingly, we find an anomalously large spectral weight transfer within a considerably broad energy range covering more than 35 eV. With an expanded energy regime of 0.5 - 35 eV we are able to accommodate important optical transitions and orbital reconfigurations corresponding to Ti, Sr and O. The ratio of the transitions between these levels changes when different defects are created and hence we are able to model the spectral weight transfer with specific defect creation.

\begin{figure}
\begin{center}
\includegraphics[width=3.4in]{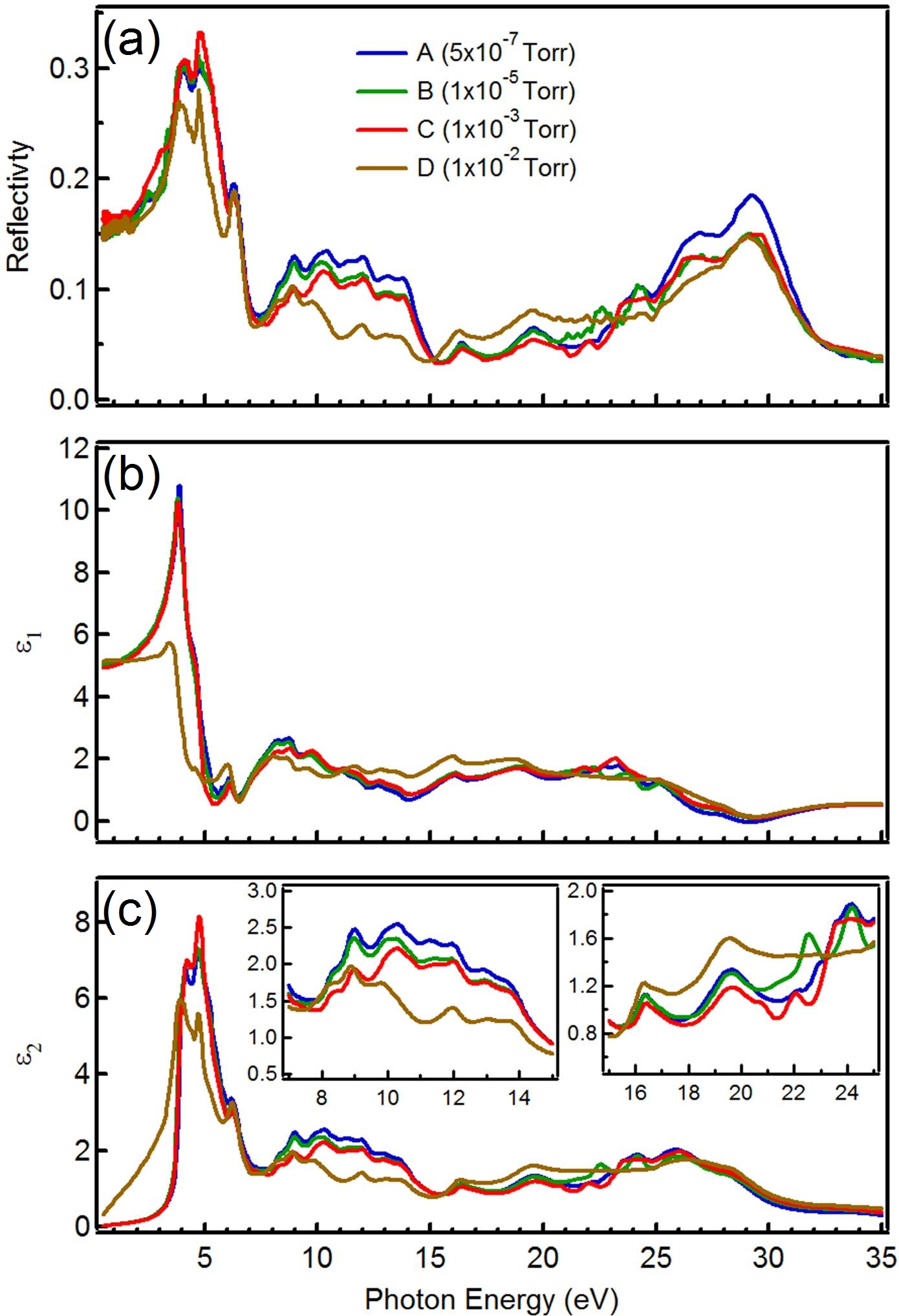}
\caption{\label{fig:fig1-R} Optical parameters of the SrTiO$_{3}$ samples measured at room temperature, (a) reflectivity, (b) real part of dielectric function, $\varepsilon_{1}$, and (c) imaginary part of dielectric function, $\varepsilon_{2}$. Insets is shown an enlarge scale of $\varepsilon_{2}$ for two different energy ranges to emphasize the intensity trend. Partial oxygen pressure ($P_{O_{2}}$) during annealing is indicated in (a).}
\end{center}
\end{figure}

Figure~\ref{fig:fig1-R} shows the room temperature reflectivity and complex dielectric function in energy of 0.5 - 35 eV for all samples (see supplementary for details\cite{supplementary}). For samples A, B and C, systematic changes occur in a broad energy range from 7.4 eV to about 35 eV, while for sample D the change is even more pronounced from below the optical band gap up to 35 eV with differences in the way spectral weight is transferred. These systematic changes in reflectivity as well as in $\varepsilon_1(\omega)$ and  $\varepsilon_2(\omega)$ within this large energy range are new observations and should play an important role for understanding the fundamental electric structure of SrTiO$_{3}$ under different $P_{O_{2}}$ anneals. A better analysis is obtained by converting the reflectivity to the $\sigma_{1}$ which is restricted by the f-sum rule, thus one can learn quantitatively the spectral weight transfer in the energy range of 0.5 to 35 eV.

\begin{figure}
\begin{center}
\includegraphics[width=3.4in]{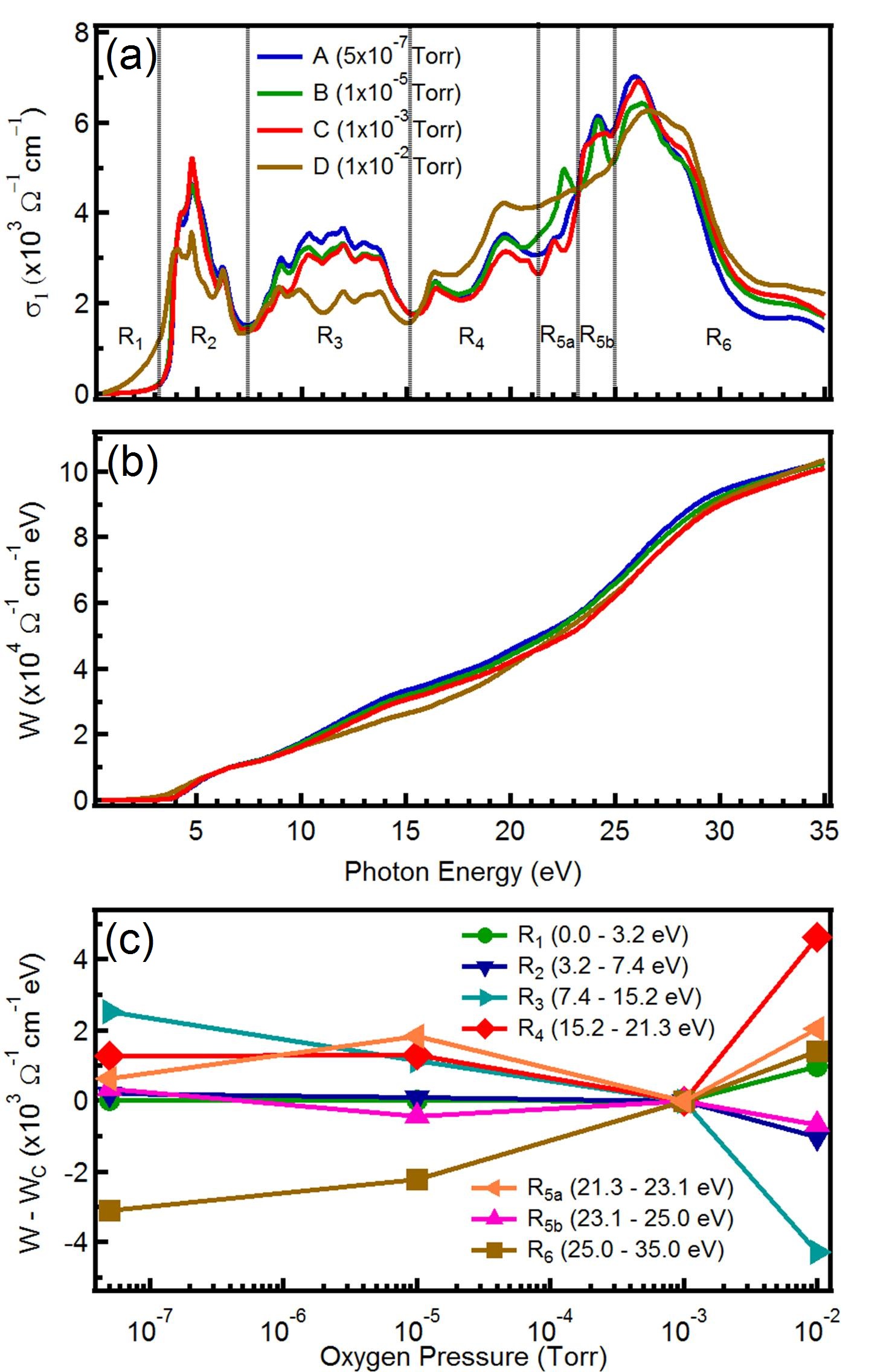}
\caption{\label{fig:fig2-S1-W} (a) Real part of optical conductivity ($\sigma_{1}$) of the SrTiO$_{3}$ samples. (b) Spectral weight $W$ of the samples plotted against the incoming photon energy. (c) $W$ extracted at various intervals of the energy range, plotted against oxygen pressure $P_{O_{2}}$ and shifted with respect to that of sample C.}
\end{center}
\end{figure}

Figure~\ref{fig:fig2-S1-W} (a) shows $\sigma_{1}$ of the samples is extracted from $\varepsilon_{2}$ by using $\sigma_{1}(\omega)=\frac{\varepsilon_{2}(\omega)\omega}{4\pi}$. For the analysis of $\sigma_{1}$, the spectrum is divided into several energy ranges. It is important to note that Cardona and van Benthem {\it et.al} have done electron energy loss spectroscopic measurements \cite{CardonaPR,vanBenthemJAP} on pristine SrTiO$_{3}$, and detail assignments are summarized in Table~\ref{table:table1trans}. The first region ($R_{1}$) is the band gap region of conventional SrTiO$_{3}$. Here, $\sigma_{1}$ of samples A, B and C is {\it flat}, i.e. there is no optically allowed transition, as expected from a band insulator. Surprisingly, the $\sigma_{1}$ of sample D shows mid-gap states with distinct and significant tail that extends all the way to very low energy, effectively closing the band gap to a large degree but not completely. This is an important signature of the cationic vacancies as discussed later.

\begin{table}
  \centering
  \begin{tabular}{|c|c|c|}
  \hline
  Region & Transition & Energy Range (eV)\\
  \hline
  $R_{1}$ & Band gap (no transition) & 0.0 - 3.2 \\
  \hline
  $R_{2}$ & Valence band $\rightarrow$ Ti-3d $t_{2g}$ & 3.2 - 7.4 \\
  \hline
  $R_{3}$ & Valence band $\rightarrow$ Ti-3d $e_{g}$, Sr-4d & 7.4 - 15.2 \\
  \hline
  $R_{4}$ & Sr-4p $\rightarrow$ Ti-3d & 15.2 - 21.3 \\
  \hline
  $R_{5a}$ & Sr-4p $\rightarrow$ Sr-4d $t_{2g}$ & 21.3 - 23.1 \\
  \hline
  $R_{5b}$ & Sr-4p $\rightarrow$ Sr-4d $e_{g}$ & 23.1 - 25.0 \\
  \hline
  $R_{6}$ & O-2s $\rightarrow$ Ti-3d, Sr-4d & 25.0 - 35.0 \\
  \hline
\end{tabular}
  \caption{\label{table:table1trans}A summary of transitions of pristine SrTiO$_{3}$. The valence band (VB) of SrTiO$_{3}$ is determined by hybridized states of (O-2p + Ti-3d $t_{2g}$) and (O-2p + Sr-4d $t_{2g}$).}
\end{table}

In the regions $R_{3}$ and $R_{4}$, the $\sigma_{1}$ value of samples A, B, and C significantly increases as $P_{O_{2}}$ decreases. The sole exception is sample D, where $\sigma_{1}$ is much lower as compared to others in $R_{2}$ and $R_{3}$, and oppositely much higher in $R_{4}$. For regions $R_{5a}$ and $R_{5b}$, the trends of $\sigma_{1}$ of sample A, B, and C are rather complex, while for sample D the $\sigma_{1}$ increases at $R_{5a}$ and decreases at $R_{5b}$. In the last region $R_{6}$, the $\sigma_{1}$ of all samples shows tendency to decrease as $P_{O_{2}}$ decreases.

An important piece of information one can extract from $\sigma_{1}$ analysis is the partial spectral weight integral ($W$) which describes the effective number of electrons excited by photons of a given energy. Further, $\sigma_{1}$ is restricted by the f-sum rule: $\int^{\infty}_{0}\sigma_{1}(E)dE=\frac{\pi n e^{2}}{2m^*}$, where $n$ is the electron density, $e$ is the elementary charge and $m^*$ is the effective electron mass. Because the number of charges involved should be conserved, the spectral weight of each sample should be eventually compensated at very high energies, and our data in Fig.~\ref{fig:fig2-S1-W} (b) clearly indicates this. The spectral weight $W\equiv\int^{E_{2}}_{E_{1}}\sigma_{1}(E)dE$ can also be extracted for finite energy ranges to extract the charge transfer between different energy regions. It is plotted in Fig.~\ref{fig:fig2-S1-W} (c) against $P_{O_{2}}$, shifted with respect to that of sample C to better emphasize the differences.

\begin{figure}
\begin{center}
\includegraphics[width=3.4in]{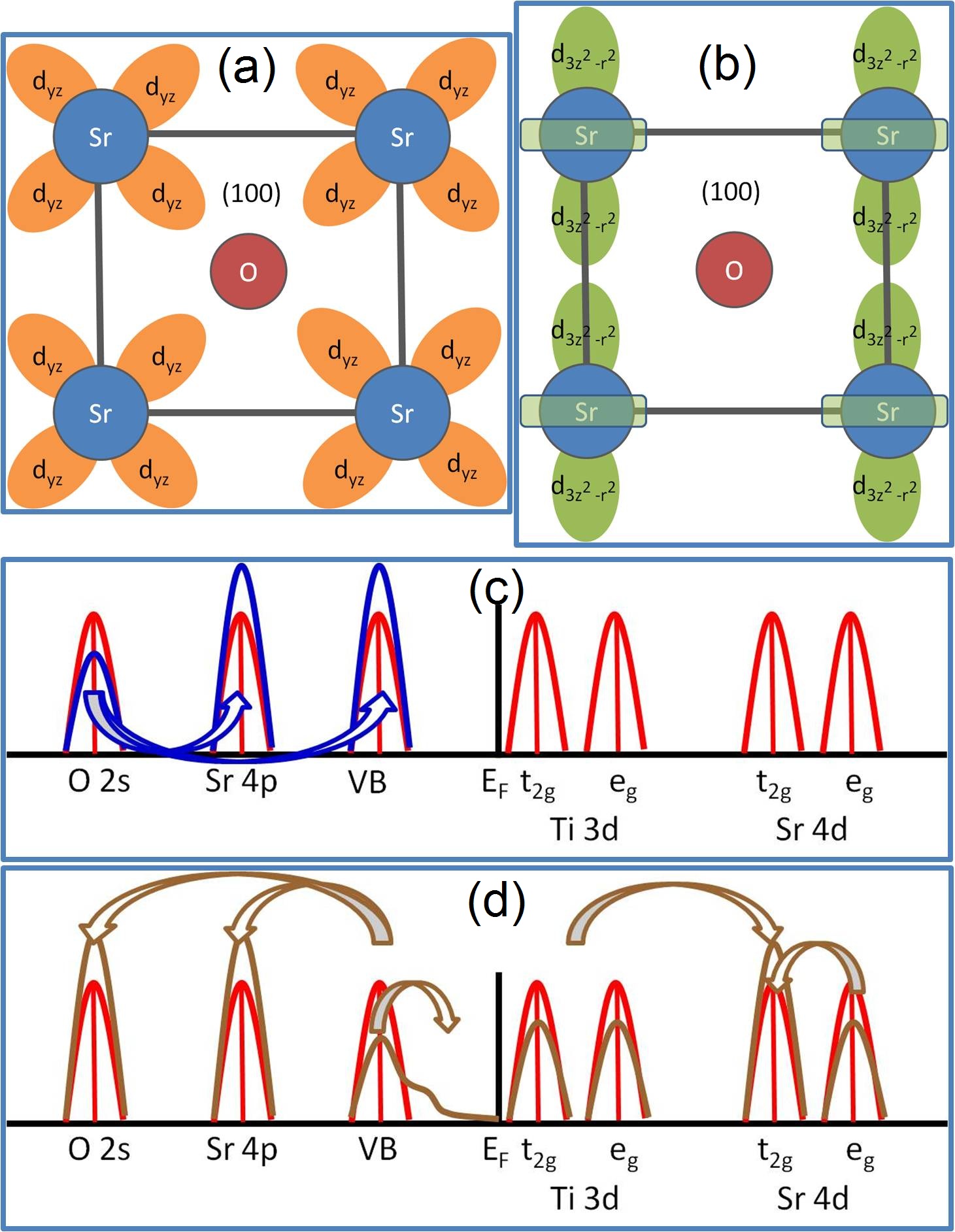}
\caption{\label{fig:fig3-model} (a) $d_{yz}$ orbital arrangement of Sr on (100) plane of SrTiO$_{3}$. (b) $d_{3z^{2}-r^{2}}$ orbital arrangement of Sr on (100) plane of SrTiO$_{3}$. (c) Charge-transfer model of SrTiO$_{3}$ at oxygen pressure below 10$^{-3}$ Torr. (d) Charge-transfer model of SrTiO$_{3}$ at oxygen pressure above 10$^{-3}$ Torr. These models are intended for qualitative illustration purpose only, and thus do not exactly follow the proper (quantitative) energy scale.}
\end{center}
\end{figure}

At $P_{O_{2}}$$<$10$^{-3}$ Torr, the decrease of $W$ at $R_{6}$, which means the decrease of O-2s occupied density of states (DOS) due to the lack of oxygen, is a direct signature of oxygen vacancies. The lack of oxygen also makes the valence state of Sr and Ti atoms to be smaller than +4, resulting in excess electrons in Ti and Sr. The excess electrons in Sr increases the availability of charges in the occupied Sr-4p states, and hence increase the $W$ of $R_{4}$, $R_{5a}$, and $R_{5b}$. A surprising observation here is the increase of $W$ at $R_{3}$. This may well be related to a signature of strong correlation effects in this system in which the increase of spectral weight near the valence band comes from much higher energy bands, i.e. O-2s (c.f. see Fig.~\ref{fig:fig3-model} and discussion below). Thus, there is a charge transfer from O-2s to Sr-4p and the valence band. Since from band structure calculation, XPS, and electron-energy-loss X-ray spectroscopy (EELS), and X-ray absorption \cite{MattheissPRB,ElliatiogluPRB,PertosaPRB,ReihlPRB,deGrootPRB,KohikiPRB} it is known that semicores O-2s states and Sr-4p states, which are hybridized with oxygen p states, are located far from the valence band ($\sim$15 to 20 eV), this signifies charge transfers across a very wide energy range, necessitating the use of high energy optics to study them properly.

At $P_{O_{2}}$$>$10$^{-3}$ Torr, the decrease of $W$ is observed at $R_{2}$ and $R_{3}$, which suggests that the availability of charges in the valence band decreases as the pressure increases. In the valence band some of the electrons are supplied by the cations, so the decrease of $W$ in these two regions indicates the presence of cationic vacancies. The presence of cationic vacancies at higher $P_{O_{2}}$ is consistent with both theoretical and experimental results \cite{MoosJACS,TanakaPRB,ZhangAdvMat,BarmanAPL}.

By examining the $W$ at $R_{4}$, which increases as $P_{O_{2}}$ increases, we find that these cationic vacancies are from Ti vacancies. As the occupied state for the transition in that region is Sr-4p states, Sr vacancies {\it cannot} be the cationic vacancies that form at high $P_{O_{2}}$. Thus, the only possible vacancies that can form are Ti vacancies. This again signifies the importance of high energy optics, since the type of the cationic vacancies can only be resolved by examining transitions at relatively high photon energies (15 - 20 eV), which otherwise would not be reached by using low energy optical techniques or transport measurements.

For the next two regions, it can be seen that at high $P_{O_{2}}$ the $W$ of $R_{5a}$ increases, while on the other hand it decreases for $R_{5b}$. If we refer to Table~\ref{table:table1trans}, this means there is a spectral weight transfer from the unoccupied Sr-4d $e_{g}$ to Sr-4d $t_{2g}$ states. The reason of this transfer can be understood if we consider the orbital configuration of the SrO plane in the SrTiO$_{3}$ perovskite lattice (Fig.~\ref{fig:fig3-model} (a) and (b)). Sr-4d $t_{2g}$ states can create stronger bonds with O since their orbital lobes ($d_{yz}$ orbital on the (100) plane in Fig.~\ref{fig:fig3-model} (a)) are closer than Sr-4d $e_{g}$ ($d_{3z^{2}-r^{2}}$ orbital on the (100) plane in Fig.~\ref{fig:fig3-model} (b)) to O atoms. Since the perovskite lattice of SrTiO$_{3}$ is symmetric in three dimensions, this also holds true for other orthogonal directions. At high $P_{O_{2}}$, the oxygen vacancy formation energy is very high, so there are very few O vacancies in the lattice as compared to the intermediate or low $P_{O_{2}}$ states. Thus, the Sr - O bond becomes stronger at high $P_{O_{2}}$, increasing the DOS of Sr-4d $t_{2g}$ since there are more oxygen atoms available to form the bond.

At high $P_{O_{2}}$, there are also more Ti vacancies in the lattice, increasing the "reliance" (and thus bond) of O to Sr due to the lack of Ti. This might also contribute to the increase of the DOS of the unoccupied Sr-4d $t_{2g}$ states. These two contributions makes the increase of the Sr-4d $t_{2g}$ DOS to be so high that it also starts to affect the $W$ of $R_{4}$. At the same time, since there are less Ti atoms, the unoccupied DOS of pure Ti-3d decreases to compensate for the increase of Sr-4d $t_{2g}$ DOS.

At $R_{1}$, a spectral tail extends all the way to very low energies as $P_{O_{2}}$ increases above 10$^{-3}$ Torr. This tail is due to an occupied mid-gap state that occurs due to the presence of Ti vacancies. This state might have O-2p characteristic, since the missing Ti atoms in the O - Ti - O bond causes the 2p orbitals of adjacent O atoms to overlap, creating a new defect state. In $R_{6}$, the $W$ increases as $P_{O_{2}}$ increases due to the simple fact that there are more O atoms available, thus increasing the DOS of O-2s. So, for $P_{O_{2}}$$>$10$^{-3}$ Torr, there is charge transfer from the valence band to the mid-gap state, Sr-4p, and O-2s, and spectral weight transfer from unoccupied Ti-3d and Sr-4d $e_{g}$ to Sr-4d $t_{2g}$. These charge transfers are depicted in Fig.~\ref{fig:fig3-model} (c) and (d).

In summary, by studying the optical conductivity over a wide range of energies up to 35 eV our result shows strong interplay between orbital reconstruction and intrinsic vacancies in SrTiO$_{3}$. Depending upon the partial pressure of oxygen during the annealing process, different types of intrinsic vacancies and orbital reconfiguration can be induced. At lower pressures, the oxygen vacancies dominate, yielding changes in the orbital configuration and spectral weight transfer from semicores O-2s to the Sr-4p and the valence band states. Contrary to this, at higher pressures Ti vacancies dominate, resulting in a spectral weight transfer from valence band states to semicores O-2s, Sr-4p as well as formation of mid-gap states and a dramatic orbital reconfiguration of both the occupied and unoccupied states. The changes of the spectral weight to such as high energy is a signature of correlations in high-energy optical conductivity of SrTiO$_{3}$. Our study shows the importance of high energy optical conductivity in understanding the intrinsic defects and the complex nature of the electronic band structure and orbital reconfiguration of SrTiO$_3$ and wide band gap oxides, in general.

We would like to acknowledge the discussion with Wei Ku. This work is supported by NRF-CRP grant Tailoring Oxide Electronics by Atomic Control, MOE Tier 2, NUS YIA, NUS cross faculty grant, FRC, Advance Material (NanoCore) R-263-000-432-646, BMBF 05KS7GUABMBF07-204, and RIKEN.

\end{document}